\documentclass[acmtog,nonacm]{acmart}

\usepackage{booktabs,amsmath} 
\usepackage{xcolor}
\usepackage{graphicx}
\usepackage{tabularx}
\usepackage{makecell}
\usepackage{bm}
\usepackage[T1]{fontenc} 

\usepackage{gensymb}

\usepackage{xcolor}

\newcommand{\Reals}{\mathbb{R}}

\newcommand{\hh}{\mathcal H}

\newcommand{\bu}{\mathbf{u}}
\newcommand{\bx}{\mathbf{x}}

\newcommand{\bi}{\bm{\omega_i}}
\newcommand{\bo}{\bm{\omega_o}}

\newcommand{\bomega}{\bm{\omega}}

\newcommand{\bv}{\bm{v}}

\newcommand{\bsigma}{\mathbf{\sigma}}
\newcommand{\bunew}{\bu_\text{new}}
\newcommand{\colorspace}{\mbox{RGB}}

\newcommand{\NOvector}{\bm{\psi}}

\newcommand{\NMBTFparams}{\bu, \bsigma, \bi, \bo}
\newcommand{\NMBTF}{M}

\newcommand{\BTF}{B}

\newcommand{\NMparams}{\bu, \bsigma}
\newcommand{\NM}{P}

\newcommand{\NTS}{T}

\newcommand{\NOMparams}{\bu, \bo}
\newcommand{\NOM}{O}

\newcommand{\NFC}{F}

\newcommand{\NOFC}{F_\text{off}}

\newcommand{\NOText}{T_\text{off}}

\newcommand{\NOfixed}{H}

\usepackage{mathtools}
\DeclarePairedDelimiter\ceil{\lceil}{\rceil}
\DeclarePairedDelimiter\floor{\lfloor}{\rfloor}

\usepackage[ruled]{algorithm2e} 

\SetAlFnt{\small}
\SetAlCapFnt{\small}
\SetAlCapNameFnt{\small}
\SetAlCapHSkip{0pt}

\newcommand\NeuMIP{NeuMIP}
\acmJournal{TOG}

\begin{document}

\setlength{\textfloatsep}{0pt}
\setlength{\abovecaptionskip}{4.0pt}
\begin{teaserfigure}
	\centering
	\setlength{\tabcolsep}{1pt}
	\begin{tabular}{c}
		\includegraphics[width=\textwidth]{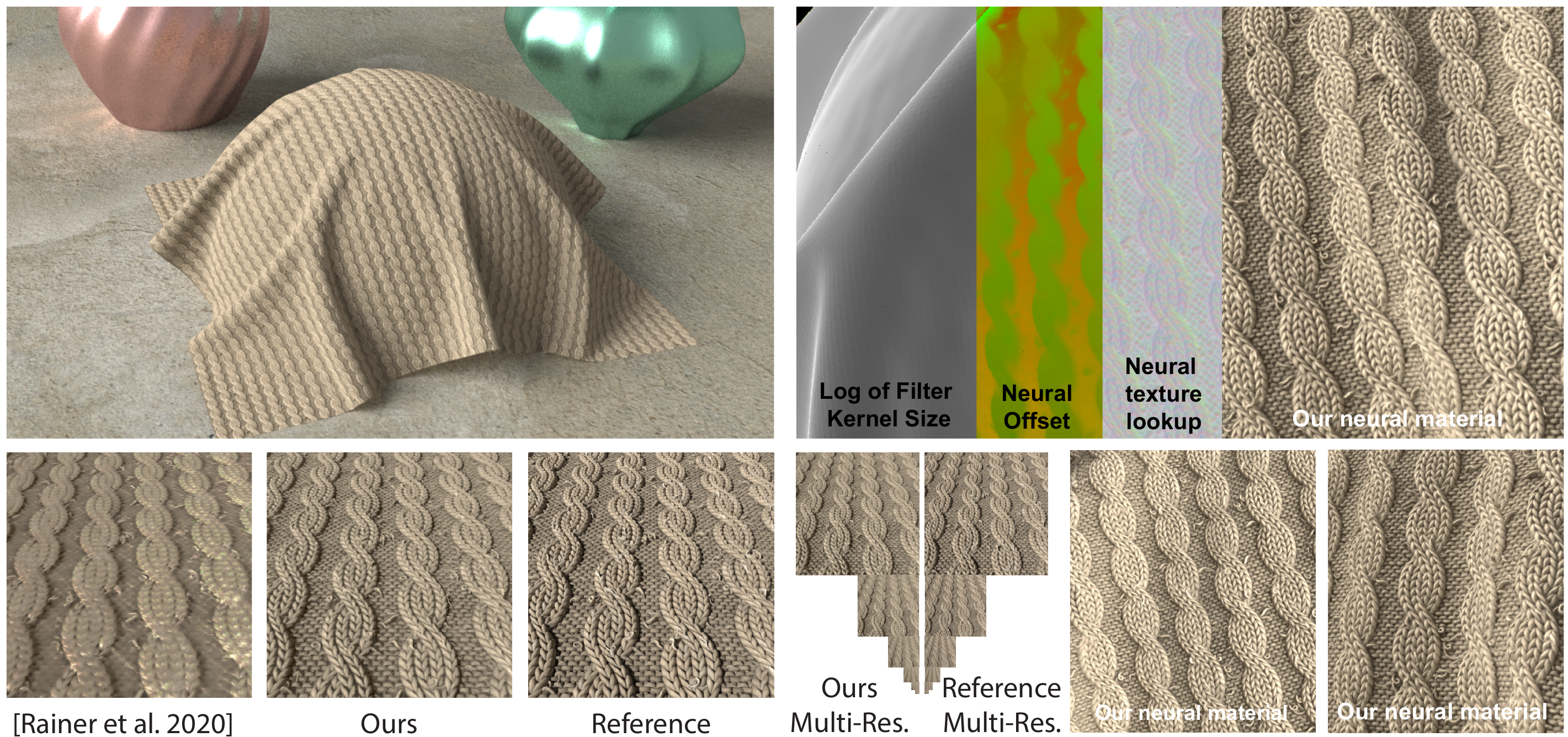} \\
	\end{tabular}
	\caption{{\bf Top left:} Our multi-resolution neural material representing Twisted Wool, rendered seamlessly among standard materials using Monte Carlo path tracing. The neural representation is trained using hundreds of reflectance queries per texel, across multiple resolutions, and is independent of the underlying input, which could be based on displaced geometry (in this example), fiber geometry, measured data, or others. {\bf Top right:} The stages of our pipeline: computing a kernel size based on pixel coverage, evaluating a neural offset module for improved handling of parallax effects, evaluating a neural texture pyramid to obtain a local feature vector, and applying a small fully-connected neural network to obtain a reflectance value usable in a standard renderer. {\bf Bottom left:} Comparison of our result to a previous technique and to a reference path-traced from the ground-truth geometry. {\bf Bottom right:} Our results match the reference across resolutions. Two additional lighting and camera angles shown.}
	\label{fig:teaser}
\end{teaserfigure}

\title{\NeuMIP: Multi-Resolution Neural Materials}


\author{Alexandr Kuznetsov}
\affiliation{%
 \institution{University of California, San Diego}
}
\email{a1kuznet@eng.ucsd.edu}

\author{Krishna Mullia}
\affiliation{%
 \institution{Adobe Research}
}
\email{mulliala@adobe.com}

\author{Zexiang Xu}
\affiliation{%
 \institution{Adobe Research}
}
\email{zexu@adobe.com}

\author{Milo\v{s} Ha\v{s}an}
\affiliation{%
 \institution{Adobe Research}
}
\email{mihasan@adobe.com}

\author{Ravi Ramamoorthi}
\affiliation{%
 \institution{University of California, San Diego}
}
\email{ravir@cs.ucsd.edu}

\begin{abstract}
We propose \NeuMIP, a neural method for representing and rendering a variety of material appearances at different scales.
Classical prefiltering (mipmapping) methods work well on simple material properties such as diffuse color, but fail to generalize to normals, self-shadowing, fibers or more complex microstructures and reflectances.
In this work, we generalize traditional mipmap pyramids to pyramids of neural textures, combined with a fully connected network. We also introduce neural offsets, a novel method which allows rendering materials with intricate parallax effects without any tessellation. This generalizes classical parallax mapping, but is trained without supervision by any explicit heightfield.
Neural materials within our system support a 7-dimensional query, including position, incoming and outgoing direction, and the desired filter kernel size. The materials have small storage (on the order of standard mipmapping except with more texture channels), and can be integrated within common Monte-Carlo path tracing systems. We demonstrate our method on a variety of materials, resulting in complex appearance across levels of detail, with accurate parallax, self-shadowing, and other effects.
\end{abstract}

%
%

%
%

\maketitle

\section{Introduction}

The world is full of materials with interesting small-scale structure: a green pasture consisting of millions of individual blades of grass, a scratched and partially rusted metallic paint on a car, a knitted sweater or a velvet dress.
The underlying mesostructure and microstructure phenomena are wildly variable: complex surface height profiles, fibers and yarns, self-shadowing, multiple reflections and refractions, and subsurface scattering. Many of these effects vary at different levels of detail: for example, we can see individual fibers of a fabric when we zoom in, but they morph into yarns and eventually disappear when we zoom out.

While computer graphics has made great strides in modeling these phenomena, this is usually at the cost of large computational expense and/or loss of generality. Many previous approaches were designed for a specific material at a particular level of detail, and evaluating those methods over a large patch becomes either slow or results in artifacts. In essence, when we zoom out, we integrate over a given patch. Theoretically, this can be achieved using Monte Carlo integration techniques by evaluating a large number of samples of the path.  However, the variance of such an estimator grows with the size of the patch, and the method quickly becomes impractical, requiring large effort to compute a function that typically becomes \emph{simpler} under zoomed-out viewing conditions. Traditional mipmap techniques \cite{williams1983pyramidal} can erroneously average parameters such as normals that influence the final appearance non-linearly. A universal method for prefiltering a material (that is, finding the integral of the patch of material microstructure covered by a pixel) has remained a challenge, despite some methods that address this problem for specific kinds of specular surfaces \cite{dupuy2013linear,jakob2014discrete} and fabrics \cite{zhao2016downsampling}.

Our goal is to develop a neural method to accurately represent a variety of complex materials at different scales, train such a method on synthetic and real data, and integrate it into a standard path-tracing system. Our neural architecture learns a continuous variant of a bidirectional texture function (BTF) \cite{dana1999reflectance}, which we term multi-scale BTF (MBTF). This is a 7-dimensional function, with two dimensions each for the query location, incoming and outgoing direction, and one extra dimension for the filter kernel size. This framework can represent a complex material (with self-shadowing, inter-reflections, displacements, fibers or other structure) at very different scales and can smoothly transition between them.

Inspired by the mipmapping technique, we propose \NeuMIP, a method that uses a set of learned power-of-2 feature textures to represent the material at different levels, combined with a fixed per-material fully connected neural network. The network takes as input the trilinearly interpolated feature vector queried from the texture pyramid, along with incoming and outgoing directions, and outputs a reflectance value. 

We also introduce a neural offset system, which allows us to efficiently represent materials with prominent non-flat geometric features. This is achieved by adjusting the texture query location through a learned offset, resulting in a parallax effect. This allows the rendering of intricate geometry without any tessellation. We obtain the appearance of a non-flat material without the cost of constructing, storing and intersecting the displaced geometry of the material.

Because our method can represent a wide variety of materials using the same architecture, adding support for a new material becomes a simple matter of creating a dataset of random material queries and optimizing the feature textures and network weights. This typically takes only about 45 minutes on a single GPU for $512^2$ resolution of the bottom pyramid level, which is easily more efficient than explicitly generating or acquiring a full high-resolution BTF. As opposed to an explicit BTF, our representation only requires storage on the order of traditional mipmapped texture pyramids (typically with 7 instead of 3 channels per texel), while enabling easy prefiltering and multiscale appearance.

Recent related works \cite{Rainer2019Neural,Rainer2020Unified} also use a neural network to efficiently compress BTFs, and we build upon these methods. However, they do not support prefiltering with arbitrary kernel sizes. They also do not have an equivalent of our neural offset technique, limiting the methods to mostly flat materials. Finally, an advantage of our method is that no encoder needs to be trained, and as a result we do not need high-resolution BTF slices as inputs, instead only requiring about 200-400 random BTF queries per texel to train a material model; our decoder network is small and fast.

The major contributions of this work are as follows:
\begin{itemize}
  \item A neural method which can represent a wide variety of geometrically complex materials at different scales, trained from random queries of the continuous 7-dimensional multiscale BTF. These queries can come from real or synthetic data.
  \item  A neural offset technique for rendering complex geometric appearance including parallax effects without tessellation, trained in an unsupervised manner.
  \item Ability to learn appearance from a small number of queries per texel (200-400), due to an encoder-less architecture.
\end{itemize}

\NeuMIP can be integrated into a Monte Carlo rendering engine, since each material query can be evaluated independently, allowing for light transport between regular and neural materials as shown in Figure \ref{fig:teaser}. 

\section{Related work}

In this section, we will briefly review previous work related to material representation and the use of deep learning in rendering.

\textbf{Prefiltering and mipmapping.} Efficiently rendering objects at different scales is one of the fundamental problems of computer graphics.
A key challenge is efficiently finding an integral of a patch of the surface of the material, which is covered by a pixel. An overview of prefiltering methods was presented by Bruneton \shortcite{bruneton2011survey}.
Williams \shortcite{williams1983pyramidal} proposed the mipmap technique to  create a pyramid of prefiltered textures; this is a standard method found in most rendering engines.
However, the prefiltering problem becomes challenging if we drop the assumption of flat and rough materials.
Many techniques were proposed over the years to address these shortcomings; however, the solutions tend to be approximate and/or focus on special cases or specific materials.
Han et al. \shortcite{Han07} uses spherical harmonics and spherical vMFs to prefilter normal maps.
Kaplanyan et al. \shortcite{kaplanyan2016filtering} proposes a real-time method for prefiltering of  normal distribution functions.
Dupuy et al. \shortcite{dupuy2013linear} prefilter displacement mapped surfaces.
Becker and Max \shortcite{becker1993smooth} introduced a method to smoothly blend between a BRDF, bump mapping, and displacement mapping. Wu et al. \shortcite{wu2019accurate} make further improvements in multi-resolution rendering of heightfield surfaces, taking into account shadowing and inter-reflection.

Parallax mapping \cite{Parallax} is a classic technique for improving bump and normal mapping by adding an approximate parallax effect. The method works by computing a texture space offset based on the local height and normal value. Our neural offset technique is inspired by this idea, but is learned unsupervised; that is, we do not feed any heightfields or normals into either the training or rendering, and in fact support materials where such heightfields/normals are not precisely defined.

\textbf{Bidirectional texture functions.} Dana et al. \shortcite{dana1999reflectance} introduced the notion of  the bidirectional texture function (BTF), a 6D function describing arbitrary reflective surface appearance. Given 2D location coordinates, incoming and outgoing directions, the BTF outputs a reflectance value. While prefiltering BTFs by mipmapping is theoretically simple, storing a discretized 6D function requires a large amount of memory. Therefore, many methods were developed to minimize storage requirements. A common solution \cite{koudelka2003acquisition, muller2003compression} is to use PCA or clustered PCA to compress the function. A comprehensive overview of different techniques was published by Filip and Haindl \shortcite{filip2008bidirectional}.

\textbf{Neural reflectance.} Recently, Rainer et al. \shortcite{Rainer2019Neural} proposed to use an autoencoder framework to compress BTF slices per texel (also termed apparent BRDFs or ABRDFs); the decoder takes incoming/outgoing directions as input in addition to the latent vector, and the autoencoder is trained per BTF. Later, they
 extended the work by unifying different materials into a shared latent space, so only a single autoencoder needs to be trained \shortcite{Rainer2020Unified}.
Within the context of complex specular appearance, \cite{kuznetsov2019learning} used Generative Adversarial Networks (GANs) to generate reflectance functions perceptually similar to synthetic or measured input data, and rendered them using partial evaluation of the generator network.
Inspired by these methods, we extend the neural textures to multi-resolution materials, and introduce the neural offset module, which greatly improves the quality of non-flat materials; we also find that neither an encoder nor a discriminator is needed in our case, and direct optimization of the feature textures works well.

\textbf{Other neural material methods.} Yan et al. \shortcite{yan2017furbssrdf} use a neural network as a mapping function to convert fur parameters into a participating media, simplifying the simulation. A neural network can also be used to accelerate the rendering in an unbiased way. Mueller et al. \shortcite{mueller2019neural, mueller20neural} presented path-guiding methods which learn a sampling distribution function for rendering using normalizing flow networks.
Nalbach et al. \shortcite{Nalbach2017b} proposed deep shading, a technique which uses a CNN to achieve screen-space effects like ambient occlusion, subsurface scattering etc, from simple feature buffers.
Thies et al. \shortcite{thies2019deferred} introduced the idea of a neural texture and using it inside deferred rendering.

Recently, Mildenhall at el. \shortcite{mildenhall2020nerf} introduced neural radiance fields (NeRFs), a view synthesis framework based on differentiable volume rendering, which can fit the geometry and outgoing radiance of 3D objects or entire scenes from a number of training views, encoding the entire appearance in a fully-connected network (MLP). This work has stimulated a large number of follow-up efforts. Our problem is simpler in that we only focus on material rather than geometric shape; however, it is more complex in other ways, as we support relighting and multi-resolution viewing, and require much faster queries to integrate in a full rendering system. We do not encode the entire reflectance in a single large MLP, and instead use more structure: a feature texture pyramid and a neural offset module combined with a much smaller MLP.

\section{Neural MBTF Representation}

\begin{table}
\begin{tabularx}{\columnwidth}{l|l|X}
{\bf Symbol} 	& {\bf Domain} 	& {\bf Definition} \\ \hline
$\hh$ 			&  				& Hemisphere \\ \hline
$\bi$ 			& $\hh$ 		& Incoming direction \\
$\bo$ 			& $\hh$ 		& Outgoing direction \\
$\bu$ 			& $\Reals^2$ 	& Surface position \\
$\bsigma$ 		& $\Reals^+ $  	& Radius of query kernel \\
$l$ 			& $\Reals $ 	& Level of detail (log. of $\bsigma$) \\
$s$ 			& $\Reals $ 	& Discrete level of detail \\
$\bv$ 			& $\Reals^c$	& Feature vector \\
$\NOvector$ 	& $\Reals^{c_2} $ & Feature v. for neural offset\\
$c,c_2$ 		& $\Reals $ 	& Feature vec. channel num. \\

\hline

$\NTS_s$ 	& $\Reals^2 \rightarrow \Reals^c$ 						&  Neural texture lookup \\
$\NMBTF$ 	& $\Reals^2,\Reals^+,\hh,\hh \rightarrow \colorspace$ 	& Multiscale BTF \\
$\NM$ 		& $\Reals^2,\Reals^+ \rightarrow \Reals^c$ 				& Neural Texture Pyramid \\
$\NFC$ 		& $\Reals^n, \hh,\hh \rightarrow \colorspace $			& Fully connected module \\
$\NOM$ 		& $\Reals^2,\hh \rightarrow \Reals^2 $ 					& Neural offset module \\
$\NOText$ 	& $\Reals^2 \rightarrow \Reals^{c_2} $  					& Neural offset texture \\

$\NOFC$ 	& $\Reals^n,\hh \rightarrow \Reals $    & Neural offset MLP \\
$\NOfixed$  & $\hh,\Reals \rightarrow \Reals^2$     & Ray depth to offset \\
\end{tabularx}
   \caption{Notation used in the paper.}
\label{tab:symbols}
\end{table}

In this section we define the multi-resolution BTF (Sec.~\ref{sec:mbtf}), and discuss our neural architecture (Sec.~\ref{sec:mntp} and \ref{sec:mno}). We introduce a baseline version of our neural MBTF in Sec.~\ref{sec:mntp} and then present our full model in Sec.~\ref{sec:mno} with a neural offset technique. In later sections, we will describe how to train our models and how to use them for rendering.
Our notation is summarized in Table \ref{tab:symbols}.

\subsection{Multi-resolution BTF} \label{sec:mbtf}

Accurately computing the reflected radiance on a material surface with complex microgeometry --- usually modeled by displaced geometry, fibers, volumetric scattering, etc. --- is highly expensive and requires tracing complex geometry at a microscopic scale.
Our approach is to precompute the reflectance of the material for a given location $\bu$, radius of the footprint kernel $\bsigma$, incoming direction $\bi$, and outgoing direction $\bo$. We denote this function $\NMBTF(\NMBTFparams)$, and call it a multi-resolution bidirectional texture function (MBTF). Below we define the MBTF more precisely.

Traditionally, a BTF \cite{dana1999reflectance} was used to incorporate all of these effects, functioning as a black-box function that directly provides the equivalent reflectance. However, a classical BTF models a material at a certain fixed scale and does not support multiple levels of detail.
We introduce the multi-resolution BTF (MBTF) that supports a continuous Gaussian query kernel size $\bsigma$, representing material appearance at an arbitrary level of detail.
Our MBTF can thus be seen as a filtered BTF.
In particular, let $G(\mu, \sigma; \bx)$ be a normalized 2D Gaussian with mean $\mu$ and standard deviation $\sigma$, as a function of $\bx$. Our MBTF is defined as follows:
\begin{equation}
	\NMBTF(\NMBTFparams) = \int_{\Reals^2} G(\bu,\bsigma;\bx) \BTF(\bu, \bi, \bo) \ d \bx,
\end{equation}
where $\BTF(\bu, \bi, \bo)$ can be seen as a traditional BTF at the finest material level. In other words, the MBTF value is the weighted average over the Gaussian kernel centered at $\bu$ of the exitant radiance in direction $\bo$, assuming the material is lit by distant light of unit irradiance from direction $\bi$.

Such a BTF can be captured from real data, which we also support; however, in most of our results, we use Monte Carlo path tracing to define it using synthetic microstructure modeled on a reference plane. We compute the value $\BTF(\bu, \bi, \bo)$ by tracing a standard path estimator from direction $-\bo$ towards $\bu$, assuming lighting from a distant light from direction $\bi$ with unit irradiance onto the reference plane. We use a distant light with a finite smoothing kernel in directions, to improve convergence of the path tracer and make multiple importance sampling applicable.

Note that our definition generalizes the standard radiometric definition of a BRDF, which is defined as outgoing radiance per unit incoming irradiance. Therefore, if the synthetic microstructure consists of a simple plane with a homogeneous BRDF, our BTF and MBTF will be equal to that BRDF for any position and kernel size.
Also note that the BTF and MBTF will in general not be reciprocal due to occlusions and multiple light bounces.



\subsection{Neural MBTF baseline} \label{sec:mntp}

\begin{figure}[tb]
\begin{center}
  \includegraphics[width=\linewidth]{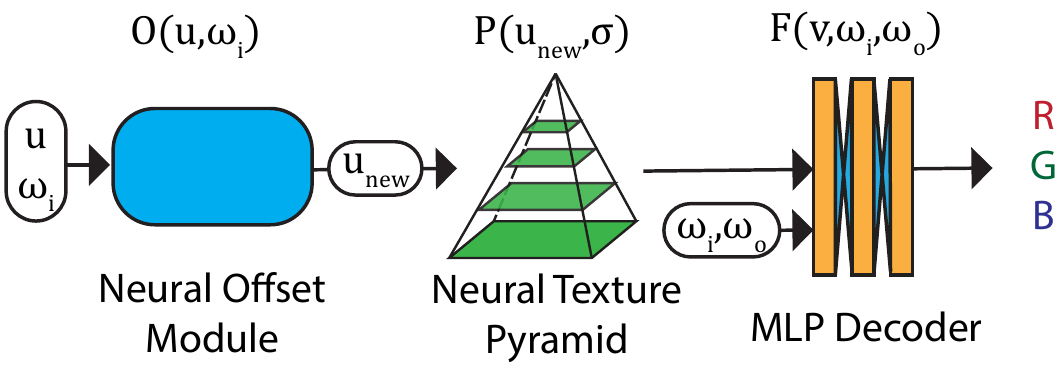}
\end{center}
    \caption{Overview of our neural architecture. {\bf Left:} The neural offset module (more detail in Fig. \ref{fig:neuraloffsetstructure}) takes a uv-space location and incoming direction, and predicts a new (offset) uv-space location to simulate parallax effects. {\bf Middle:} The neural texture pyramid is queried using trilinear interpolation to obtain a 7-channel feature vector. {\bf Right:} The feature vector, with incoming and outgoing directions, is fed to a material-specific multi-layer perceptron, which predicts the RGB reflectance value.}
\label{fig:overview}
\end{figure}

The 7-dimensional MBTF is too prohibitive to store directly.
Nevertheless, this function (depending on the material) has low entropy because of redundancies, which can be captured by a well-chosen neural network architecture.
We propose a baseline neural architecture that can already model many complex materials.
In particular, our baseline architecture consists of a neural texture pyramid $\NM$ that encodes complex multiscale spatially-varying material appearance and a material decoder network $\NFC$ that computes directional reflectance from the pyramid.
This neural MBTF is expressed by
\begin{equation}
    \NMBTF(\NMBTFparams) = \NFC(\NM(\NMparams), \bi, \bo),
    \label{eq:baseline}
\end{equation}
where $\NM(\NMparams)$ represents a neural feature lookup from the neural pyramid $\NM$ and the decoder network $\NFC$
evaluates the final reflectance from the neural feature given input directions ($\bi$ and $\bo$).

\paragraph{Neural texture pyramid}
Instead of having one single neural texture of dimension $2^k \times 2^k \times c$ (where $k$ is some positive integer and $c$ the number of channels), we leverage a neural texture pyramid $\NM = \{\NTS_s\}$, consisting of a set of neural textures $\NTS_s$.
Each $\NTS_s$ has a size $2^s \times 2^s \times c$, where each texel contains a $c$-channel latent neural feature, $s$ denotes the discrete level of details, and we let $0 \le s \le k$.
Such a pyramid structure enables encoding complex material appearance at multiple scales.
This is similar to standard mipmapping for color (or BRDF) textures; however, our neural pyramid models challenging appearance effects caused by complex microgeometry.
In addition, unlike traditional mipmaps, our per-level neural textures are independent from one another: there is no simple operation that produces all texture levels from the finest one. Each neural texture in the set is independently optimized, which ensures that the MBTF can represent appearance for the material at all levels.

We utilize trilinear interpolation to fetch a neural feature $\NM(\NMparams)$ at location $\bu$ and (continuous) level $\bsigma$.
Much like in classical mipmapping, we use standard bilinear interpolation to get features for the spatial dimensions of the location coordinate $\bu$, followed by linear interpolation in the logarithmic space for the scale dimension, since the scales express powers of 2.
Specifically, a neural feature $\bv$ at $(\bu, \bsigma)$ is computed by:
\begin{equation}
    \bv = \NM(\NMparams) = w_1\NTS_{\floor{l}}(\bu) + w_2\NTS_{\ceil{l}}(\bu),
    \label{eq:nm}
    \end{equation}
where $l=\log_2(\bsigma)$, $w_1=(\ceil{l}-l)$, $w_2=(l-\floor{l})$ and $\floor{l},\ceil{l}$ are floor and ceiling operations. $\NTS_s(\bu)$ is the bilinearly-interpolated neural texture lookup from the level $s$ with resolution $2^s\times2^s$ with infinite tiling (wrap-around).

\paragraph{Material decoder}
We design our material decoder $\NFC$ as a multi-layer perceptron (MLP) network to regress the final reflectance from the  neural feature queried from the neural pyramid.
The input of this network contains $c+4$ values, consisting of $c$ values from neural textures and 4 values from the directions (encoded as 2D points on the projected hemisphere).
One of the design goals of our architecture is fast evaluation.
Therefore, our multi-layer perceptron consists of only 4 layers. Each intermediate layer has 25 output channels.
The final 3-channel output is used as the RGB reflectance value of the material.
We use ReLU as the activation function for all layers, including the final layer where it clamps the final output to be non-negative.
Note that, unlike \cite{thies2019deferred} that uses a CNN to process neural textures for global context reasoning in scene rendering,
our goal is modeling realistic material with locally complex microgeometry.
Therefore, we enforce that our neural textures express the spatial information, encoding complex microgeometry effects;
thus a light-weight MLP is able to efficiently decode the texture for our task and is also fast to evaluate.

\subsection{Neural MBTF with Neural Offset} \label{sec:mno}

\begin{figure}[tb]
\begin{center}
    \includegraphics[width=\linewidth]{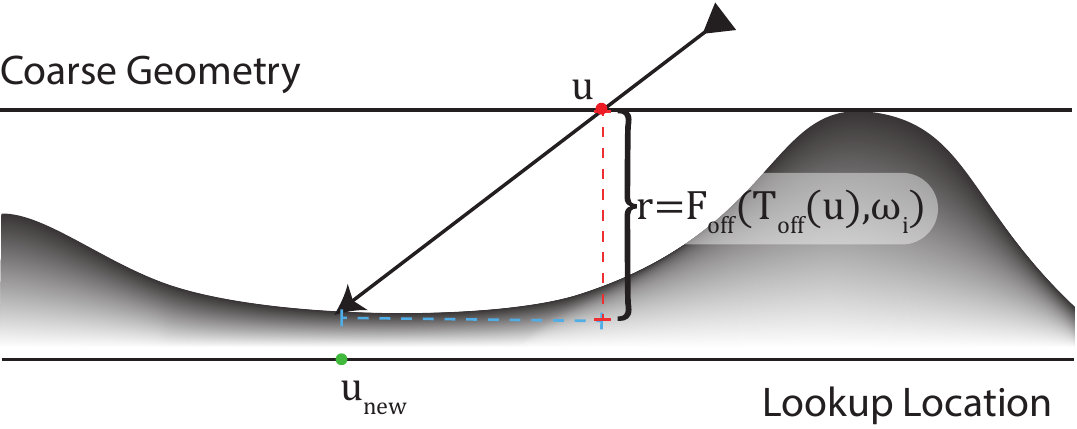}
\end{center}
    \caption{An intuition for the fixed function stage of the neural offset module. Instead of predicting the uv-space offset directly, we predict the depth under the intersection with the reference plane, and convert it to the offset by assuming a locally flat surface. Note however that no actual heightfield is used to supervise the method, and for some materials (e.g. volumetric or made from fibers) such a heightfield is not even precisely defined.}
\label{fig:neuraloffseintersection}
\end{figure}

\begin{figure}[tb]
\begin{center}
    \includegraphics[width=\linewidth]{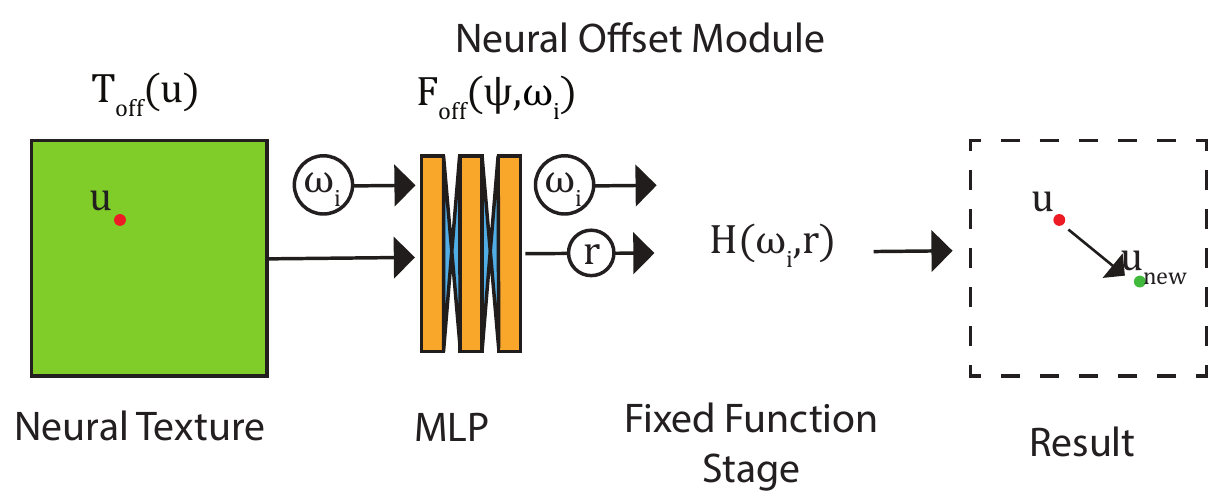}
\end{center}
    \caption{Details of the neural offset module, which is shown in blue in Fig. \ref{fig:overview}. A bilinear query of the neural offset texture results in a feature vector, which (together with incoming direction) is fed into a multi-layer perceptron. The resulting scalar value is interpreted as a ray offset, which is converted into a 2D offset in texture space by the fixed function stage.}
\label{fig:neuraloffsetstructure}
\end{figure}

\begin{figure*}[tb]
\begin{center}
    \includegraphics[width=\textwidth]{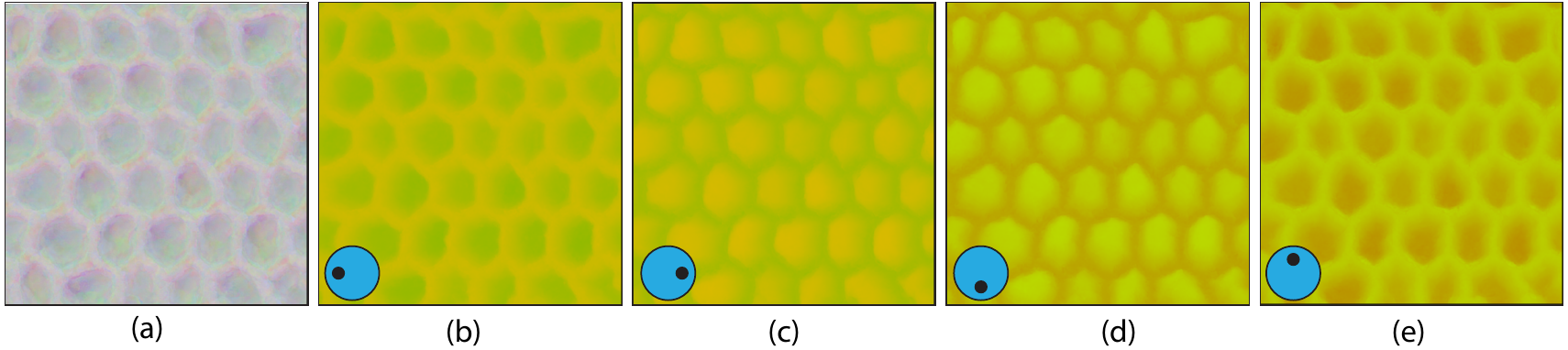}
\end{center}
    \caption{Neural offset visualization on the \emph{turtle shell} material (which is shown in Section \ref{sec:results} and the supplementary video). {\bf (a)} The learned neural offset feature texture (3 out of 7 channels shown as RGB colors). {\bf (b-e)} The computed 2D offset, color-coded using red/green for the two components. The blue circle shows the incoming ray direction on a projected hemisphere.}
\label{fig:neural_offset_vis}
\end{figure*}

Our baseline neural MBTF can already model many real materials;
however, it is very challenging for this network to handle highly non-flat materials that have significant parallax and occlusion effects.
Although, by increasing its capacity, a big enough neural network can potentially approximate any function, this is not ideal, as bigger networks lead to longer rendering times; they also result in a slower training rate as the neural texture needs to learn correlated information across multiple pixels for different camera directions. We have also observed poor angular stability when learning non-flat materials in a generic way.


To improve results on complex non-flat materials, we introduce a neural offset module: a network that predicts a coordinate offset for the feature texture lookup. Instead of directly using the intersection location $\bu\in \Reals^2$, we use the network module $\NOM(\NOMparams)$ to calculate a new lookup location:
\begin{align}
\bunew = \bu + \NOM(\NOMparams)
\end{align}
This is shown in Fig. \ref{fig:neuraloffseintersection}. With the help of neural offsets, we can slightly adjust the lookup location of the texture depending on the viewing direction, achieving effects such as parallax.

\paragraph{Neural offset}
Instead of directly regressing a 2D offset, we train a network to regress a 1D scalar value $r$ representing the ray depth at the intersection which can be easily turned into the final 2D offset given the view direction $\bo$ (see Fig.~\ref{fig:neuraloffseintersection}).
This makes our model more geometric-aware, easing the neural offset regression task.
In particular, the neural offset module consists of 3 components: a neural offset texture $\NOText$, an MLP $\NOFC$ that regresses the ray depth $r$ from $\NOText$, and a final step $\NOfixed$ (a fixed function) that outputs the offset (See Fig. \ref{fig:neuraloffsetstructure}).
The design of using a neural texture and an MLP is similar to our baseline MBTF network described above, except the texture look-up is just bilinear (no pyramid).
Specifically, the ray depth $r$ is computed by
\begin{align}
r=\NOFC(\NOvector, \bo)=\NOFC(\NOText(\bu), \bo),
\end{align}
where $\NOvector = \NOText(\bu)$ is the latent feature vector lookup in $\NOText$ on the initial location $\bu$.
The MLP $\NOFC$ takes the latent vector $\NOvector$ and the viewing direction $\bo$ as input; it again consists of 4 layers, and each layer outputs 25 channels (except for the last one), with ReLU activation functions in between.
Given the estimated ray depth $r$, the offset is computed by
\begin{align}
    \NOfixed(r, \bo) = \frac{r}{\bomega_z}(\bomega_x,\bomega_y),
\end{align}
where $\bomega_x,\bomega_y,\bomega_z$ are the components of $\bo$. Therefore, the final form of the neural offset query is:
\begin{equation} \label{eq:nomFull}
\bunew = \bu + \NOfixed(r, \bo) = \bu + \frac{r}{\bomega_z}(\bomega_x,\bomega_y).
\end{equation}
The new lookup location $\bunew$ can now be used in place of $\bu$ to lookup a latent vector in the neural texture pyramid in eq. \ref{eq:baseline}. Note that the network can also learn the 2D offset function $\NOM(\NOMparams)$ directly, but in our experiments the result was worse without the constraint of a hard-coded function $\NOfixed(r, \bomega)$. Figure \ref{fig:neural_offset_vis} visualizes the predicted offset learned by the module on a highly non-flat material.

\paragraph{Full neural MBTF representation}
Our full neural representation is modeled by prepending the the neural offset module to our baseline neural MBTF network.
Basically, we use a neural offset module to get a new location $\bunew$ (Eq. \ref{eq:nomFull}) by translating the original input $\bu$.
Then we use that $\bunew$ to query a feature vector from the neural texture pyramid $\NM$ (eq. \ref{eq:nm}). Finally, we use an MLP material decoder $\NFC$, in conjunction with incoming/outgoing directions, to get the final reflectance value. Our full neural MBTF can be described as follows:
\begin{equation}
    \NMBTF(\NMBTFparams) = \NFC(\NM(\bu+\NOM(\NOMparams), \bsigma), \bi, \bo).
    \label{eq:NMBTF}
\end{equation}
This whole framework is fully differentiable, enabling end-to-end training that simultaneously optimizes the neural offset, the multi-scale pyramid, and the decoder.
We train the neural offset module in an unsupervised way. We do not provide any ground truth offsets;
in fact, for some materials (e.g. volumetric or fiber-based) there may be no clear surface and therefore no well-defined correct offset.
The end-to-end training allows our full model to jointly auto-adjust the multiple neural components and leads to the best visual quality it can achieve.
Please refer to section \ref{sec:training} for the training procedure.




\begin{figure*}[htp]
\begin{center}
    \includegraphics[width=160mm]{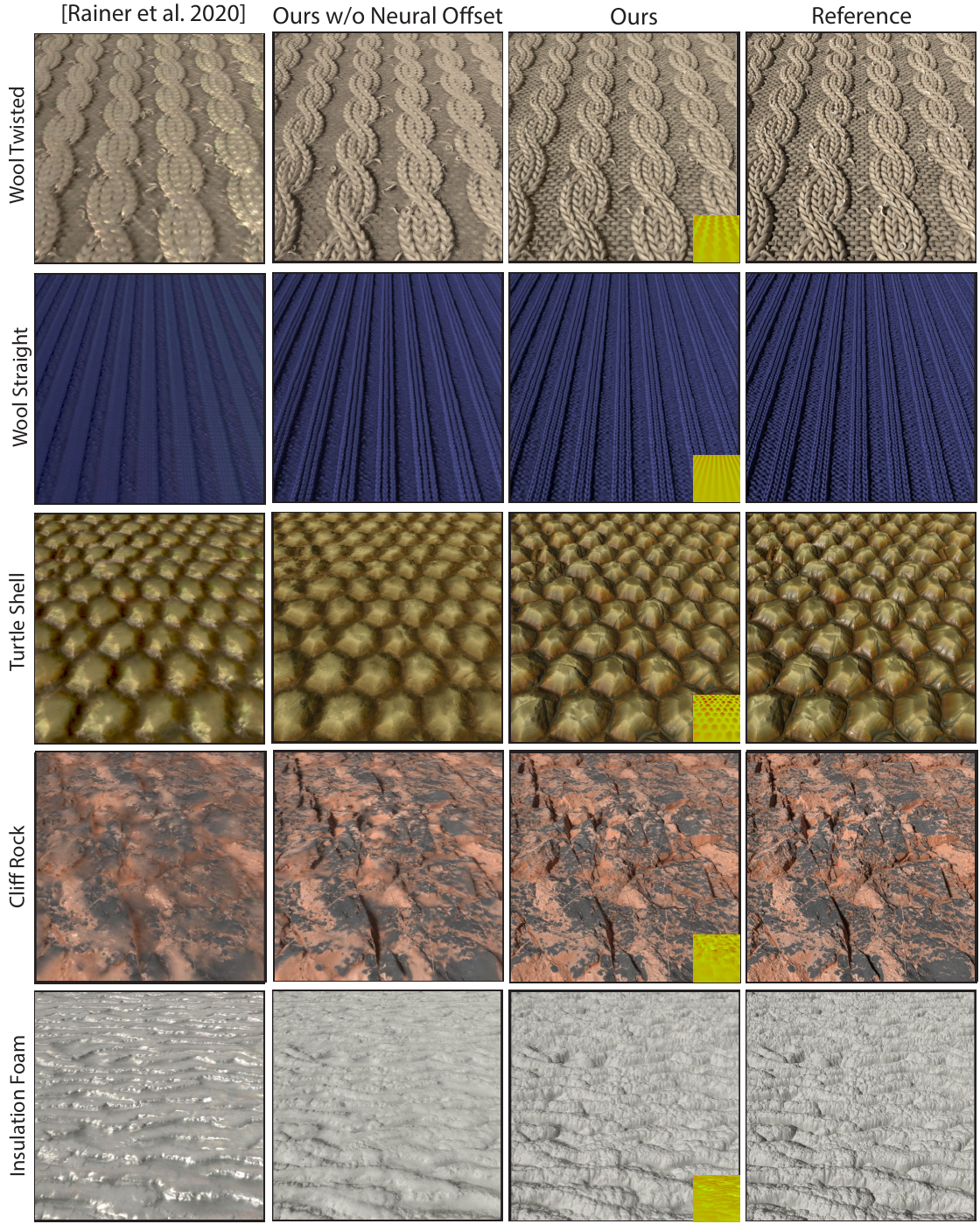}
\end{center}
    \caption{{\bf Comparison.} We show a comparison of several different methods: Rainer et al. \shortcite{Rainer2020Unified}, our baseline method (without neural offset), our full method, and reference computed by path-tracing of synthetic material microstructure. The materials are mapped to a plane, viewed at an angle and lit by a single light slightly to the left. Our baseline method already outperforms Rainer et al., despite being trained with fewer BTF queries. Our neural offset adds even better handling of parallax effects. The small insets in ``Ours'' show the color-coded 2D neural offset. The match with reference is close, with a minor loss in shadow contrast (the hard shadows form a reflectance discontinuity which is hard to learn perfectly).}
\label{fig:results_flat_plane}
\end{figure*}

\section{Data Generation and Training} \label{sec:training}

\paragraph{Synthetic data preparation} We generate synthetic MBTF data by first constructing the microgeometry, and using a CPU-based standard path tracer with a custom camera ray generator to render the required queries; we use smoothed directional lighting with unit irradiance on the reference plane, to ensure valid radiometric properties. In most cases, the microgeometry is constructed by meshing and displacing a high-resolution heightfield texture, and driving other reflectance properties (albedo, roughness, metallicity, micro-scale normal) from additional textures. The synthetic MBTF data is generated in about 30 minutes on a 32-core CPU, using 64 samples per query in a commercial rendering engine. In the basket weave example in Fig. \ref{fig:results_mitsuba1}, we use a fiber-level representation without any meshed surface, applying the fiber shading model of Chiang et al. \shortcite{Chiang2016}; this material is precomputed using the PBRT renderer \cite{PBRT}.

\paragraph{Training} Our neural module is fully differentiable and can be trained end-to-end.
As the input to the module we provide 7D queries consisting of the light direction, camera direction, uv-location and kernel radius. The network produces RGB colors for the given queries in a forward pass, and back-propagation updates the network weights and neural textures.
One training batch consists of around a million queries ($2^{20}$); this number is much larger than the number of input MBTF queries per texel, so one batch updates many texels. We train the network until convergence (typically 30000 iterations). The training time is about 45 minutes for a $512^2$ maximum resolution, and about 90 minutes for a $1024^2$ maximum resolution, using a single NVIDIA RTX 2080 Ti GPU.

If we optimize neural feature vectors individually, this can result in noisy neural textures. As a result, objects rendered with such materials will have a noisy appearance, reminiscent of Monte Carlo noise.
This is especially true for the neural offset texture.
We have developed a technique to avoid those problems. During training, we apply a Gaussian blur with initial standard deviation of $\sigma_i=8$ texels to the neural textures. As the training progresses, we relax $\sigma$ exponentially over time with a half life $h=3333$ iterations: $\sigma(t)=\sigma_i \cdot 2^{-t/h}$.




\begin{table}[tb]
\setlength{\tabcolsep}{2pt}
\footnotesize
\begin{tabular}{r||c|c|c|c||c|c|c|c}
 & \multicolumn{4}{c}{ MSE $\times10^{-3}$} & \multicolumn{4}{c}{ LPIPS} \\ \hline
 & \shortstack{Our \\ w/o \\ n.o.} & Our & Rainer  & \shortstack{Rainer \\ /our \\ ratio} & \shortstack{Our \\ w/o \\ n.o.}  & Our & Rainer   &\shortstack{Rainer \\ /our \\ ratio}  \\
 \midrule
Wool Twisted & 11.31 & \textbf{3.98} & 19.03 & 4.78 & 0.333 & \textbf{0.151} & 0.501 & 3.32 \\
Wool Straight & 1.04 & \textbf{0.39} & 1.92 & 4.93 & 0.236 & \textbf{0.112} & 0.403 & 3.59 \\
Turtle Shell & 3.35 & \textbf{1.31} & 4.84 & 3.70 & 0.527 & \textbf{0.161} & 0.477 & 2.96 \\
Cliff Rock & 6.00 & \textbf{2.00} & 10.26 & 5.12 & 0.316 & \textbf{0.142} & 0.473 & 3.35 \\
Insulation Foam & 9.18 &\textbf{ 3.90} & 21.75 & 5.57 & 0.425 & \textbf{0.196} & 0.561 & 2.87 \\
\end{tabular}
\caption{Errors for images in Fig. \ref{fig:results_flat_plane}. Note that both our per-pixel MSE and perceptual LPIPS scores are consistently better.}
\label{tab:results}
\end{table}

\section{Rendering}

Our neural materials can be integrated into Monte Carlo rendering engines, so that light can seamlessly interact across regular objects and objects with neural materials. We implemented our final rendering in the Mitsuba rendering engine \cite{Mitsuba}. If a surface with our neural material is hit, we need to evaluate the neural module. We also need to sample an outgoing direction for indirect illumination.  For simplicity, we sample indirect rays according to the cosine-weighted hemisphere, which is sufficient for our current examples. Note that for each shading point, we need to evaluate the material up to twice: for the light sample and for the BRDF sample.


There are multiple choices for implementing our neural module in a practical rendering system. We could certainly port its implementation to C++, since the neural network is a simple 4-layer MLP, which just requires 4 dense matrix multiplications. However, our current solution is to reuse the PyTorch code to ensure exactly matching outputs between training and rendering. Because the heavy-lifting operations are anyway implemented in C++/CUDA via the PyTorch framework, the Python overhead is negligible. This required some modifications to the Mitsuba path integrator. We use material query buffers: in our integrator, we trace a path until we encounter a neural material. When that happens, we put the query in the buffer, and continue tracing. When the buffer is full or if there are no more active paths to trace, we send the buffer to PyTorch/GPU for evaluation. 


\section{Results} \label{sec:results}

\begin{table}[t]
\setlength{\tabcolsep}{2pt}
\begin{tabular}{r||c|c|c|c|c|c|c|c}
\footnotesize Material  LoD & 0 & 1 & 2 & 3 & 4 & 5 & 6 & 7 \\
\midrule
\footnotesize Wool Twisted & 3.98 & 2.10 & 1.47 & 1.07 & 0.74 & 1.16 & 1.17 & 1.08 \\
\footnotesize Wool Straight & 0.39 & 0.16 & 0.11 & 0.08 & 0.09 & 0.05 & 0.03 & 0.10 \\
\footnotesize Turtle Shell & 1.31 & 0.92 & 0.70 & 0.67 & 0.88 & 0.82 & 0.55 & 0.48 \\
\footnotesize Cliff Rock & 2.00 & 0.87 & 0.63 & 0.51 & 0.42 & 0.32 & 0.24 & 0.10 \\
\footnotesize Insulation Foam & 3.90 & 1.64 & 1.20 & 0.86 & 0.76 & 0.55 & 0.57 & 0.49 \\
\end{tabular}
\caption{MSE Errors (scaled $\times10^{-3}$)  for images rendered across multiple levels of detail, shown in Fig. \ref{fig:results_lod}. Note that the errors are generally becoming \emph{lower} for more distant zooms, i.e. larger filter kernels.}
\label{tab:results_lod}
\end{table}

\begin{table}[t]
\begin{tabular}{r||l|l|l}
 & Ours & Rainer & Ratio \\
 \midrule
\# of network weights & \textbf{3332} & 38269 & 11.5 \\
\# of texture channels & \textbf{14} & 38 & 2.7 \\
\end{tabular}
\caption{The total number of network parameters. For fair comparison, we only count the number of weights in the decoder of Rainer et al.'s method, as the encoder is not needed for deployment in a rendering system. We use 7 channels per texel for both the neural offset texture and the feature texture pyramid.}
\label{tab:parameters}
\end{table}

\begin{figure*}[tb]
\begin{center}
    \includegraphics[width=\textwidth]{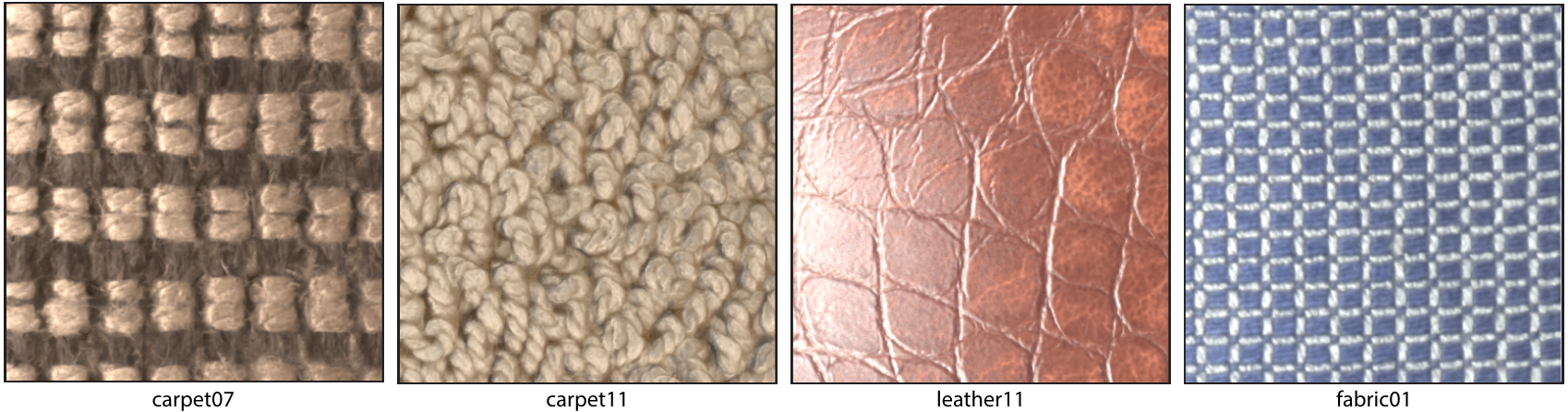}
\end{center}
    \caption{{\bf Real BTF data.} We also observe good results when applying our method to real BTF data acquired from physical material samples. Our method is trained with a random set of BTF queries, and these can come from any source.} 
\label{fig:btf_results}
\end{figure*}

In this section, we showcase the abilities of our neural method to represent and render a range of material appearances, with training data coming from different input representations: displaced heightfield geometry with varying reflectance properties, fiber geometry, and measured data.

\paragraph{Comparisons} In Figure \ref{fig:results_flat_plane}, we show results rendered on a flat plane, with camera and directional light at an angle. We compare several different methods: Rainer et al. \shortcite{Rainer2020Unified}, our baseline method (without neural offset), our full method (with neural offset), and a ground truth computed by path-tracing of the synthetic material structure. The materials are mapped to a plane, viewed at an angle and lit by a single light slightly to the left. Our baseline method already outperforms the universal encoder of Rainer et al., despite being trained with fewer BTF queries. We believe this is due to our decoder-only architecture, which can adapt to the material and benefits from a stochastic distribution of the input BTF queries, and our improved training techniques (especially the progressively decaying spatial Gaussian blur). The multi-resolution nature of our solution also helps. On the other hand, Rainer et al.'s solution has the benefit of very fast encoding in case the queries are already in the required uniform format, and its performance could likely improve if the encoder was retrained on a different distribution of materials that matches our examples more closely.

Our neural offset adds even better handling of parallax effects on top of our baseline result. The small insets in ``Ours'' show the color-coded 2D neural offset. The match with reference is close, with a minor loss in shadow contrast (the hard shadows form a reflectance discontinuity which is hard to learn perfectly).

\paragraph{Real BTF results} While we mostly focus on synthetically generated BTF queries, we also support fitting real BTF data (acquired from physical material samples and also used by Rainer et al.) using our neural architecture. This is demonstrated by Figure \ref{fig:btf_results}. Since our method is trained with a random set of BTF queries, these can come from any source.

\paragraph{Quantitative evaluation} We also measure the numerical error for images in Fig. \ref{fig:results_flat_plane} when compared to the reference, both using a per-pixel MSE (Mean Squared Error) and perceptual LPIPS (Learned Perceptual Image Patch Similarity) scores. Our scores, shown in Table \ref{tab:results}, are consistently better than Rainer et al.'s.

\begin{figure}[t]
\begin{center}
    \includegraphics[width=\columnwidth]{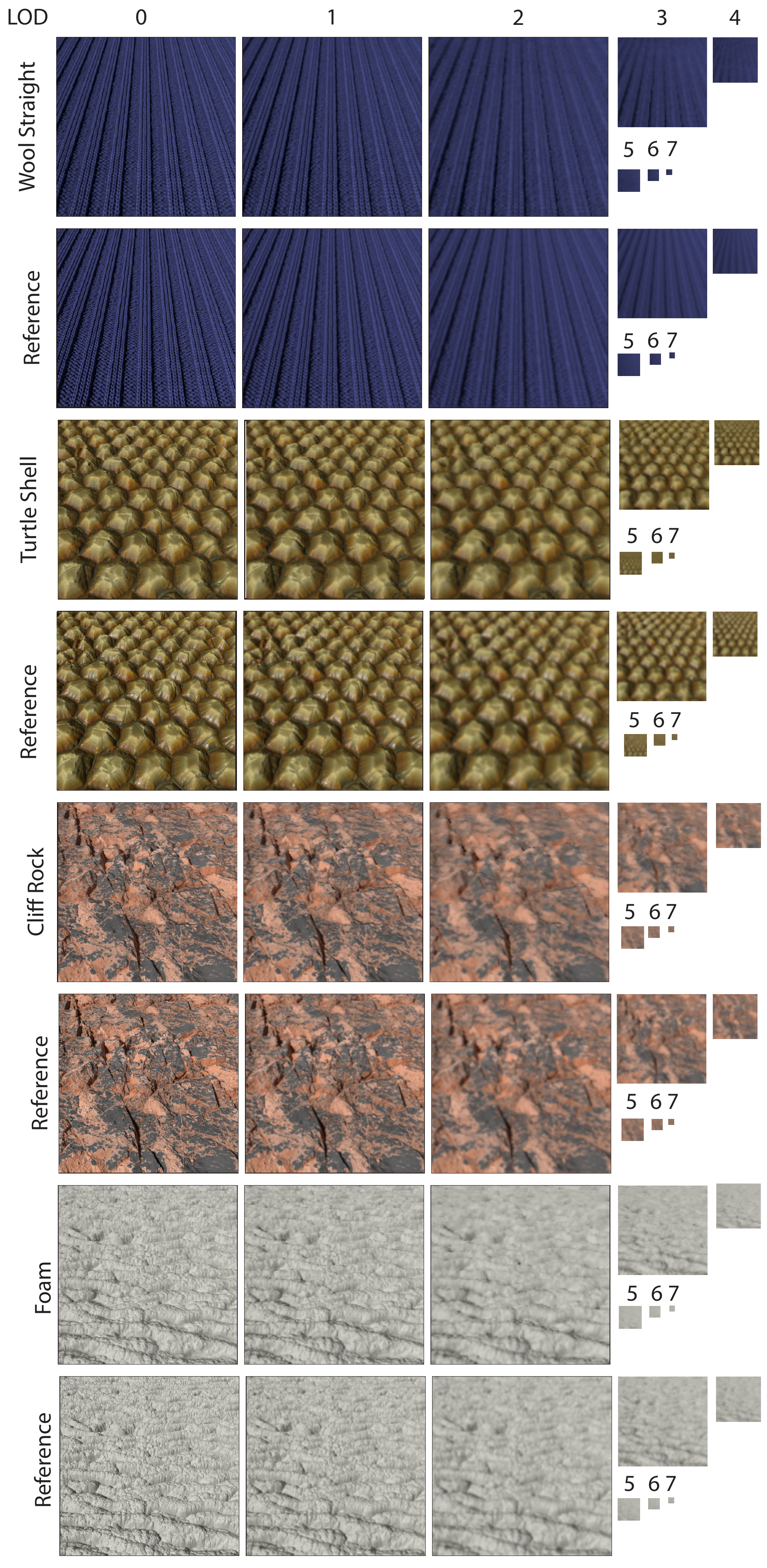}
\end{center}
    \caption{Materials rendered at the different level of details. Because the camera is at a 45\degree, the actual LOD of the image is slightly higher at the tops of the images and slightly lower at the bottoms of the images. }
\label{fig:results_lod}
\end{figure}

\paragraph{Multiresolution results}
In Table \ref{tab:results_lod}, we also report the MSE scores for images rendered across multiple levels of detail, shown visually in Fig. \ref{fig:results_lod}. We observed that with higher (coarser) levels in the multiresolution hierarchy, the errors actually tend to decrease. This is because materials at a coarse level of detail tend to have fewer high-frequency details. As a result, it becomes even easier to optimize corresponding feature vectors from the neural texture pyramid and the network at those levels of detail.

\paragraph{Network Size}
We also compare our neural network sizes to Rainer et al.'s in Table \ref{tab:parameters}. Not only does our method have smaller errors, it has 11.5 times fewer weight parameters and uses less than half the number of channels for the neural textures compared to Rainer et al. This is because the task of  Rainer et al.'s decoder is significantly harder. It needs to decode a latent vector which was encoded by a universal (and necessarily imperfect) encoder. In contrast, the task of our network (decoder) is to decode a more specialized latent vector, which was created through direct optimization for a specific material. For this reason, our network requires substantially fewer weights. By dividing our network into two stages via a neural offset module, we increase the network representative power compared to a simple MLP architecture.

\begin{figure*}[p]
\begin{center}
    \includegraphics[width=\textwidth]{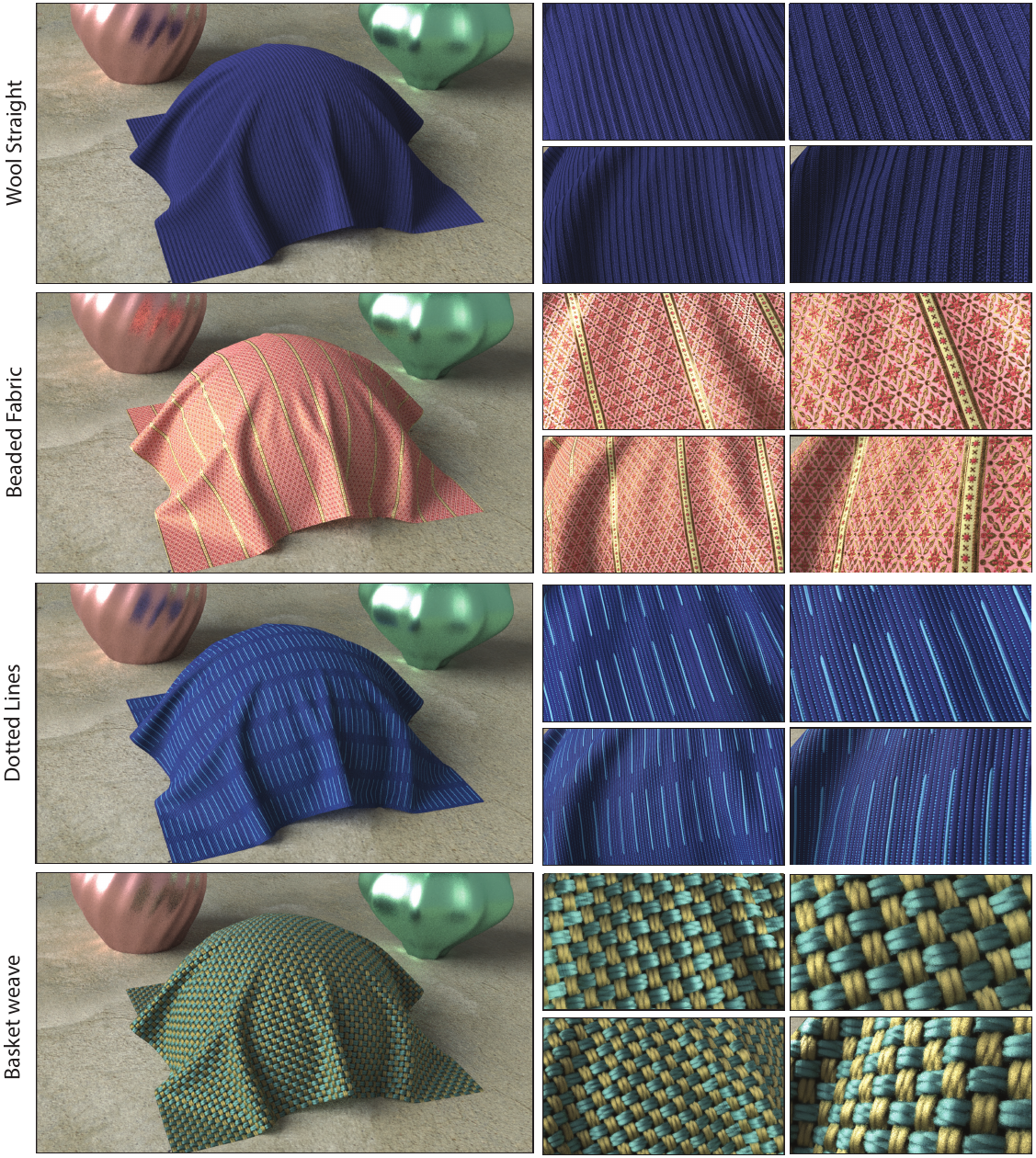}
\end{center}
    \caption{{\bf Fabric} materials rendered with our method. The fabrics are modeled as heightfields with spatially varying reflectance properties, except the last Basket Weave example, which is constructed from yarns using actual fiber geometry shaded with the model of Chiang et al. \shortcite{Chiang2016} and does not use a heightfield. We show several camera, light directions, and multiple zoom levels. Please see the supplementary video for extensive animated versions.}
\label{fig:results_mitsuba1}
\end{figure*}

\begin{figure*}[p]
\begin{center}
    \includegraphics[width=\textwidth]{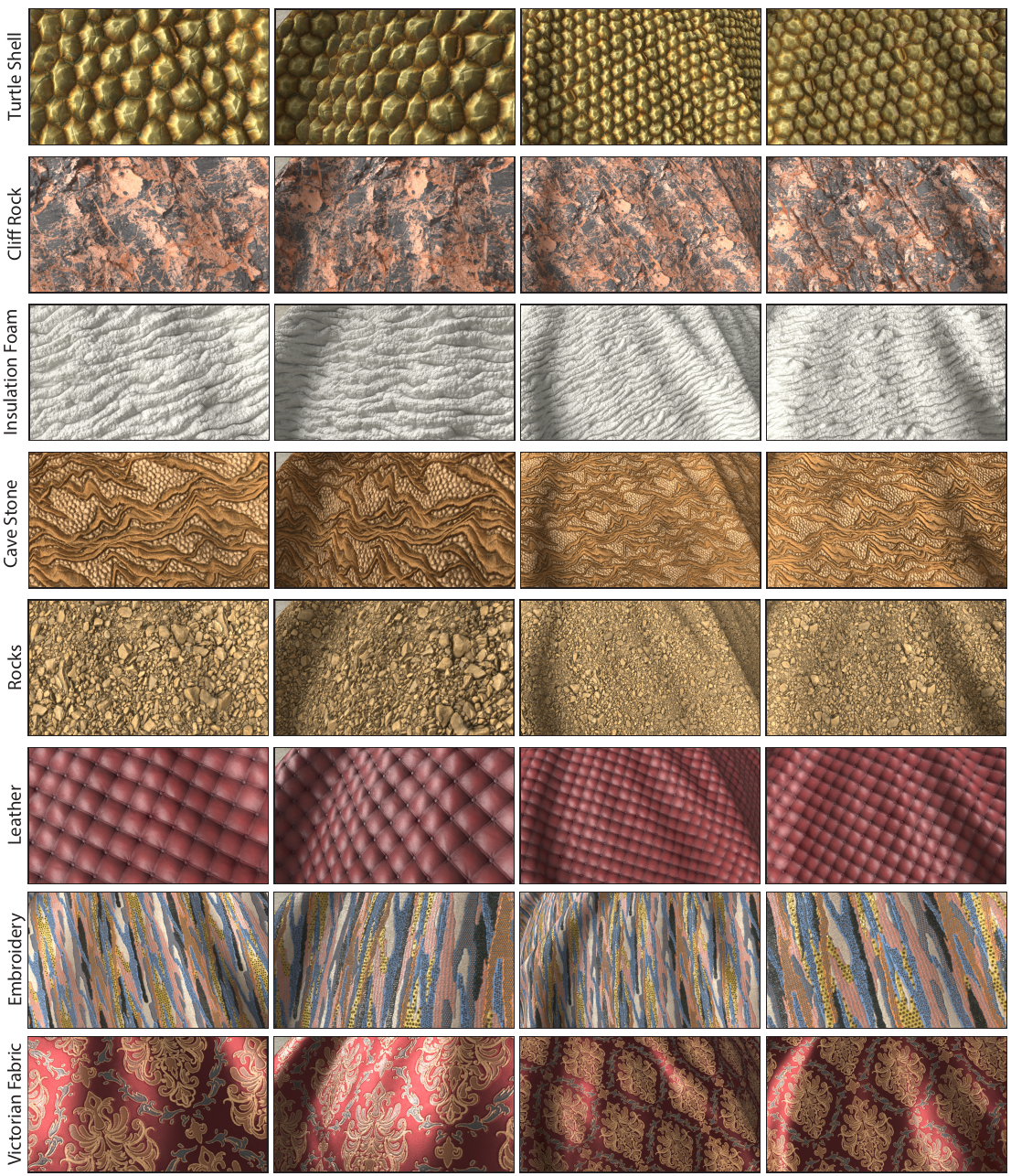}
\end{center}
    \caption{An assortment of other materials rendered with our method, showing different camera views and light directions. Please make sure to view their animated versions in the supplementary video, which is important to fully appreciate parallax/occlusion effects.}
\label{fig:results_mitsuba2}
\end{figure*}

\paragraph{Additional renderings} In Figure \ref{fig:results_mitsuba1}, we show several fabric-like materials rendered at multiple levels of zoom. The top three fabrics are modeled as heightfields with spatially varying reflectance properties. The textures driving the height and reflectance are from the Substance Source library \cite{Subs}, though any textures can be used. The last Basket Weave example is constructed from yarns using actual fiber geometry shaded with the model of Chiang et al. \shortcite{Chiang2016} and does not use a heightfield. We show several camera and light directions in the right column. In Figure \ref{fig:results_mitsuba2}, we show a further selection of materials rendered with our method, showing different camera views and light directions. Please make sure to view their animated versions in the supplementary video, which is important to fully appreciate parallax/occlusion effects.

\paragraph{Limitations} One limitation of our current implementation is simple importance sampling. For more specular materials, this could be extended by fitting the parameters of a parametric pdf model (e.g. a weighted combination of a Lambertian and microfacet lobe of a given roughness) per texel. The benefit of this solution would be that no additional neural network is required to sample, and path continuations can be decided independent of network inference. Another limitation is that very specular (e.g. glinty) materials would be hard to handle without blurring; this could be addressed by inserting randomness into our decoder, to synthesize some specular detail instead of attempting to fit it exactly.

\section{Conclusion and Future Work}

We presented a neural architecture that can be trained to accurately represent a variety of complex materials at different scales. Our neural architecture learns a multi-scale bidirectional texture function (MBTF): a 7-dimensional function, with two dimensions each for the query location, incoming and outgoing direction, and one dimension for the filter kernel radius. As part of the architecture, we introduced a neural offset technique for rendering complex geometric appearance including parallax effects without tessellation, trained in an unsupervised manner. Our encoder-less architecture can be trained from a small number of random queries per texel (200-400). These queries can come from real or synthetic data. We show a number of results, demonstrating high quality appearance with accurate displacement, parallax, self-shadowing, and other effects. We believe this approach will stimulate further research on neural representations of materials that are difficult or expensive to handle with classical methods.

The most exciting future work avenue, in our opinion, is to fully explore the set of material structures that can be procedurally generated as inputs to our method. Other interesting directions include more advanced importance sampling (e.g. by fitting multi-lobe parametric distributions per texel); the parameters of such a distribution could be stored in small additional textures. A straight-forward extension would be to support semitransparent materials by predicting alpha transparency in addition to reflectance. Yet another direction would be to make the method support glinty specular effects, perhaps by inserting an additional random vector into the decoder, and training it with a GAN loss to generate stochastic detail matching the input data distribution.

\section{Acknowledgments}

We thank Xuezheng Wang for an interactive OptiX port. We thank Sitian Chen and Steve Marschner for the fiber-level fabric microstructure data. This work was funded in part by NSF grant 1703957 and NSF Chase-CI, ONR DURIP N000141912293, the Ronald L. Graham Chair and the UC San Diego Center for Visual Computing.

\bibliographystyle{ACM-Reference-Format}
\bibliography{references}

\end{document}